\documentstyle[12pt]{article}
\begin{document}
\begin{center}
{\LARGE\bf Analytic loops and gauge fields}
\footnote{Nucl. Phys. B606 (2001) 636.}
\bigskip
\par
{\bf E.K. Loginov}\footnote{e-mail: loginov@ivanovo.ac.ru}
\par
{\sl Physics Department, Ivanovo State University
\par
Ermaka St. 39, Ivanovo, 153025, Russia}
\end{center}
\begin{abstract}
In this paper the linear representations of analytic Moufang loops are investigated. We prove that every representation of semisimple analytic Moufang loop is completely reducible and find all nonassociative irreducible representations. We show that such representations are closely associated with the (anti-)self-dual Yang-Mills equations in ${\bf R}^8$.
\end{abstract}

\section{Introduction}
The pure Yang-Mills theory defined in the four-dimensional Euclidean space has a rich and interesting structure even at the classical level. The discovery of regular solutions to the Yang-Mills field equations, which correspond to absolute minimum of the action (Belavin et al.)~[1], has led to an intensive study of such a classical theory. One hopes that a deep understanding of the classical theory will be invaluable when one tries to quantize such a theory.
\par
In the past few years, increased attention has been paid to gauge field equations in space-time of dimension greater than four, with a view to obtaining physically interesting theories via dimensional reduction~[2]. A telling illustration of this is the geometrical Higgs mechanism due to Englert and others~[3]. At the same time, Corrigan et al.~[4] have obtained an analog of (anti-)self-dual Yang-Mills equations in ${\bf R}^8$. Later there were found its several solutions~[5,6] which were then used to construct string and membrane solitons~[6,7]. In recent papers~[8], the 8D equations have been applied to construct a topological Yang-Mills theory on Joyce manifolds as an 8D counterpart of the Donaldson-Witten theory~[9]. It is also recently discussed that self-dual Yang-Mills gauge fields depending only upon time play a role in the context of M-theory~[10].
\par
From the viewpoint of mathematical physics, the above works has made most conspicuous the possibly central role played by octonions~[11] and their attending Lie groups~[12]. The algebra of octonions (Cayley numbers) is the most known example of nonassociative alternative algebras. The alternative algebras are closely associated with the Malcev algebras and analytic Moufang loops. These algebraic structures are actively investigated and applied in physics~[13]. Our work is a step in this direction.
\par
The paper is organized as follows. Section 2 contains well known facts of the structural theory of alternate algebras, Malcev algebras, and analytic Moufang loops which we use. Sections 3 and 4 present our main results. In Section 3 the theorem on connection of analytic Moufang loop representations and its tangent Malcev algebra representations is proved. The complete reducibility of semisimple analytic Moufang loop representations is proved and all irreducible nonassociative representations of analytic Moufang loops are found. In Section 4 we show that such representations are closely associated with the (anti-)self-dual Yang-Mills equations in ${\bf R}^8$ and find solutions of the latter.

\section{Analytic Moufang loops}
A loop is a binary system $G$ with an unity element 1, in which the equations $ax=b$ and $ya=b$ are uniquely solvable for all $a,b\in G$. Moufang loops are distinguished from the class of all loops by the identities
\begin{eqnarray}
ab\cdot ca=a(bc\cdot a),\quad (cb\cdot c)a=c(b\cdot ca),\quad a(b\cdot cb)=(ab\cdot c)b
\end{eqnarray}
which are called the central, left, and right Moufang identity accordingly. It is well known~[14], that two of them are a corollary of third. This paper is concerned with analytic Moufang loops; that is, analytic manifolds equipped with the Moufang loop structure, in which the binary operations are analytic.
\par
Moufang loops are closely associated with the alternative algebras which are defined by the identities
\begin{eqnarray}
x^{2}y=x(xy),\qquad yx^{2}=(yx)x.
\end{eqnarray}
It is evident from the definition that any associative algebra is alternative. We shall now a construction of the most important alternative algebras which are not associative, namely, the Cayley-Dickson algebras[15,16].
\par
Let $A$ be an algebra with an involution $x\to\bar x$ over a field $F$ (here and below we consider finite-dimensional algebras of characteristic 0 with identity element). Given a nonzero $\alpha\in F$ we define a multiplication on the vector space $(A,\alpha)=A\oplus A$ by
\begin{eqnarray}
(x_1,y_1)(x_2,y_2)=(x_1x_2-\alpha \bar y_2y_1,y_2x_1+y_1\bar x_2).
\end{eqnarray}
This makes $(A,\alpha)$ an algebra over $F$. It is clear that $A$ is isomorphically embedded into $(A,\alpha)$ and $dim(A,\alpha)=2dimA$. Let $e=(0,1)$. Then $e^2=-\alpha$ end $(A,\alpha)=A\oplus Ae$. Given any $z=x+ye$ in $(A,\alpha)$ we suppose $\bar z=\bar x-ye$. Then the mapping $z\to\bar z$ is an involution in $(A,\alpha)$.
\par
Starting with the base field $F$ the Cayley-Dickson construction leads to the following tower of alternative algebras:
\par
1. $F$, the base field.
\par
2. ${\bf C}(\alpha)=(F,\alpha)$, a field if $x^2+\alpha$ is the irreducible polynomial over $F$; otherwise, ${\bf C}(\alpha)\simeq F\oplus F$.
\par
3. ${\bf H}(\alpha,\beta)=({\bf C}(\alpha),\beta)$, a generalized quaternion algebra. This algebra is associative but not commutative.
\par
4. ${\bf O}(\alpha,\beta,\gamma)=({\bf H}(\alpha,\beta),\gamma)$, a Cayley-Dickson algebra. Since this algebra is not associative the Cayley-Dickson construction ends here.
\par
The algebras in 1 --- 4 are called composition. Any of them has the nondegenerate quadratic form (norm) $n(x)=x\bar x$, such that $n(xy)=n(x)n(y)$. In particular, over the field ${\bf R}$ of real
numbers, the above construction gives 3 split algebras (if $\alpha=\beta=\gamma=-1$) and 4 division algebras (if $\alpha=\beta=\gamma=1$): the fields of real ${\bf R}$ and complex ${\bf C}$ numbers, the algebras of quaternions ${\bf H}$ and octonions ${\bf O}$, taken with the Euclidean norm $n(x)=\mid x\mid$. Note also that any simple nonassociative alternative algebra is isomorphic to the Cayley-Dickson algebra ${\bf O}(\alpha,\beta,\gamma)$~[16].
\par
The set of all regular elements of alternative algebra $A$ over ${\bf R}$ is an analytic Moufang loop. Its tangent algebra is isomorphic to the commutator algebra $A^{(-)}$ of $A$. If $A$ is nonassociative algebra, then $A^{(-)}$ is not a Lie algebra. Instead of the Jacobi identity, $A^{(-)}$ satisfies the Malcev identity
\begin{eqnarray}
J(x,y,xz)=J(x,y,z)x,
\end{eqnarray}
where $J(x,y,z)\equiv (xy)z+(yz)x+(zx)y$ is the Jacobian of $x,y,z$, and the Sagle identity
\begin{eqnarray}
(xy\cdot z)t+(yz\cdot t)x+(zt\cdot x)y+(tx\cdot y)z=(xz)(yt).
\end{eqnarray}
The identities (2.4) and (2.5) are equivalent~[17]. An anticommutative algebra whose multiplication satisfies these identities is called a Malcev algebra.
\par
Malcev algebras and alternative algebras are closely associated. Any simple non-Lie Malcev algebra is isomorphic to the commutator algebra of Cayley-Dickson algebra~[18,19]. In particular, there exists a
unique simple compact not-Lie Malcev algebra over ${\bf R}$. It is isomorphic to the seven-dimensional algebra ${\bf O}^{(-)}$. Any semisimple Malcev algebra $A$ is decomposed in the direct sum $A=N(A)\oplus J(A)$ of Lie center $N(A)$ and ideal $J(A)$. Besides, $N(A)$  is a semisimple Lie algebra and $J(A)$ is a direct sum of simple non-Lie Malcev algebras~[19,20].
\par
There exists a correspondence between local analytic Moufang loops and real Malcev algebras~[21]. It generalizes the classical correspondence between local Lie groups and Lie algebras. The
correspondence is completely transferred for (global) analytic Moufang loops~[22]. Namely, there exists an unique simply connected analytical Moufang loop $G$ with the given  tangent Malcev algebra, and any connected analytic Moufang loop with the same tangent algebra is isomorphic to the quotient algebra $G/H$ where $H$ is a discrete central normal subgroup in $G$. Any simply connected semisimple analytic Moufang loop is decomposed in a direct product of semisimple Lie group and simple nonassociative Moufang loops each of which is analytically isomorphic to one of the spaces ${\bf S}^7$, ${\bf S}^3\times{\bf R}^4$, or ${\bf S}^7\times{\bf R}^7$. Actually, any simply connected simple nonassociative Moufang loop is isomorphic to the loop of elements of norm 1 in the Cayley-Dickson algebra over ${\bf R}$ or ${\bf C}$.

\section{Bimodules and birepresentations}
We recall the concepts of bimodule and birepresentation of linear alge\-bra~[23]. Let $\cal K$ be a class of linear algebras over a field $F$. Suppose we have a pair of linear mappings $(\rho,\lambda):A\to EndM$ of $A\in\cal K$ into a vector space $M$ over $F$. Then we can define a multiplication on the vector space direct sum $A\oplus M$ by
\begin{eqnarray}
(a,x)(b,y)=(ab,x\rho_{b}+y\lambda_{a}).
\end{eqnarray}
It is clear that this operation is bilinear, so $A\oplus M$ is an algebra. Moreover, $A$ is a subalgebra and $M$ is an ideal in $A\oplus M$ such that $M^{2}=0$. The algebra $A\oplus M$ is called the split null extension of $A$. If $A\oplus M\in\cal K$, then $M$ is called a bimodule for $A$ (or $A$-module) and the pair of mappings $(\rho,\lambda)$ a birepresentation of $A$ in the class $\cal K$.
\par
It is comfortable to use the designations $xa$ and $ax$ instead of $x\rho_{a}$ and $x\lambda_{a}$. Clearly, we can consider the algebra $A$ as $A$-bimodule if $xa$ and $ax$ are the
multiplication in $A$. Such bimodules and corresponding birepresentation $a\to R_{a}$, $a\to L_{a}$ are called regular.
\par
A bimodule $M$ for an alternate algebra $A$ is alternative $A$-module if and only if
\begin{eqnarray}
(a,b,x)=-(x,b,a)=-(a,x,b)=(b,x,a)
\end{eqnarray}
where $a,b\in A$, $x\in M$ and $(a,b,x)=ab\cdot x-a\cdot bx$. Examples of alternative bimodules are the regular bimodule for an alternative algebra and the Cayley-Dickson bimodule for an associative composition algebra. The latter is the $A$-submodule $Ae$ of the composition algebra $(A,\alpha)=A\oplus Ae$ where $\rho$ and $\lambda$ are induced by the left and right multiplications of $A$ on $(A,\alpha)$. There is a second way of representing the Cayley-Dickson bimodule: to taking $M=A$, $\rho_{a}$ and $\lambda_{a}$ the right multiplications in $A$ by $\bar a$ and $a$ respectively.
\par
It is known~[24], that every alternative bimodule for a semisimple alternative algebra is completely reducible. The irreducible bimodule for an alternative algebra is either associative or a submodule of the regular bimodule for the Cayley-Dickson algebra or a submodule of the Cayley-Dickson bimodule for the generalized quaternions algebra.
\par
By (2.5) the linear mapping $\tau:A\to EndM$ is called a (right) representation of Malcev algebra $A$ if
\begin{eqnarray}
\tau_{xy\cdot z}=\tau_{x}\tau_{y}\tau_{z}-\tau_{z}\tau_{x}\tau_{y}+\tau_{y}\tau_{zx}-\tau_{yz}\tau_{x}.
\end{eqnarray}
for all $x,y,z\in A$. In this case $M$ is called a Malcev $A$-module. Since the algebra $A$ is anticommutative, the concepts of Malcev $A$-module and bimodule are equivalent. An special case of the Malcev representation is the regular representation $x\to R_{x}$.
\par
It is known~[19,25], that every representation of semisimple Malcev algebra is completely reducible. The irreducible bimodule is either Lie or the regular bimodule for a simple nonassociative Malcev
algebra or $sl(2,F)$-module of dimension 2 such that $\tau (a)=\bar a$ where $\bar a$ is the adjoint matrix to the matrix $a\in sl(2,F)$. Since every semisimple Malcev algebra $A$ can be embedded in the commutator algebra of appropriate semisimple alternative algebra, the representation $\tau$ of $A$ is connected to the representation $(\rho,\lambda)$ of the alternative algebra by the equality $\tau=\rho-\lambda$.
\par
The concepts of bimodule and birepresentation of linear algebra can be extend to Moufang loops. Let $M$ be a linear space over a field $F$ and $(\rho_{a},\lambda_{a})$ invertible linear transformations of $M$ where $a\in G$. Define the multiplication (3.1) on the set $G\times M$. If the groupoid $G\times M$ is a Moufang loop, then we say that the vector space $M$ is $G$-module and the pair of mappings $(\rho,\lambda):G\to AutM$ a linear representation of $G$. If $G$ is an analytical Moufang loop, we additionally require $G\times M$ to be also analytic loop.
\par
Suppose $F[G]$ is the formal linear envelope of the Moufang loop $G$ and $M$ is $G$-module. Extend the representation $(\rho,\lambda):G\to AutM$ from $G$ to $F[G]$ by $F$-linearity. We'll have the representation $(\tilde\rho,\tilde\lambda):F[G]\to EndM$ of $F[G]$. The sets
\begin{eqnarray}
Ker (\rho,\lambda)&=&\{a\in G\mid\rho_{a}=\lambda_{a}=id\},
\nonumber\\
Ker(\tilde\rho,\tilde\lambda)&=&\{a\in F[G]\mid\tilde\rho_{a}=\tilde\lambda_{a}=0\}
\nonumber
\end{eqnarray}
is called kernels of the representation $(\rho,\lambda)$ and $(\tilde\rho,\tilde\lambda)$ accordingly. The representation $(\rho,\lambda)$ is called faithful if its kernel coincides with the unity element of $G$. In the previous paper~[26], we proved the following
\par\medskip
{\bf Proposition 1.} {\sl The mappings $(\rho,\lambda):G\to AutM$ are a linear representation of the Moufang loop $G$ if and only if:
\par
(i) The associator $(a,b,x)$ is skew-symmetric.
\par
(ii) $x(b\cdot ab)=(xb\cdot a)b$ and $(ab\cdot a)x=a(b\cdot ax)$
\par\noindent
for all $a,b\in F[G]$ and $x\in M$}.
\par\medskip
{\bf Proposition 2.} {\sl The following statements are true:
\par
(i) $Ker(\rho,\lambda)$ is a normal subloop of $G$.
\par
(ii) $Ker(\tilde\rho,\tilde\lambda)$ is an ideal in $F[G]$.
\par
(iii) The quotient algebra $F[G]/Ker(\tilde\rho,\tilde\lambda)$ is alternative.}
\par\medskip
{\bf Proposition 3.} {\sl Every irreducible $G$-module of Moufang loop $G$ is a submodule of an irreducible alternative bimodule.}
\par\medskip
Now let $G$ be a subloop of the loop of invertible elements of an alternative algebra $A$. Mark out the subalgebra $B$ in $A$ generated by elements of $G$. The operators $R_{a}$ and $L_{a}$ of the right and left multiplications by the element $a\in G$ are invertible, satisfy Proposition 1, and coincide with the identity operator only if $a=e$ where $e$ is unity element of $G$. Therefore the pair of mappings $(R,L):G\to AutB$ is a faithful representation of $G$.
\par
Conversely, let the Moufang loop $G$ has the faithful representation $(\rho,\lambda):G\to AutM$. Extend it from $G$ to $F[G]$ by $F$-linearity. The homomorphism $\tilde\varphi:F[G]\to
F[G]/Ker(\tilde\rho,\tilde\lambda)$ of algebras induces the homomorphism $\varphi:G\to G/H$ of loops. We'll prove that the subloop $H$ coincides with the kernel $Ker(\rho,\lambda)$ of the
representations $(\rho,\lambda)$.
\par
Let $g\in Ker(\rho,\lambda)$. Then $e-g\in Ker(\tilde\rho,\tilde\lambda)$ and $\varphi(e)-\varphi(g)=\varphi(e-g)=0$. Therefore $g\in H$ and $Ker(\rho,\lambda)\subseteq H$. Let $g\in H$. Then $\varphi(e-g)=\varphi(e)-\varphi(g)=0$ end $e-g\in Ker(\tilde\rho,\tilde\lambda)$. Therefore
$g\in Ker(\rho,\lambda)$ and $H\subseteq Ker(\rho,\lambda)$. Thus, $H=Ker(\rho,\lambda)$.
\par
Since the representation $(\rho,\lambda)$ is faithful, the loop $G$ is isomorphically embedded into the loop of invertible elements of $F[G]/Ker(\tilde\rho,\tilde\lambda)$. Using Proposition 2, we get
\par\medskip
{\bf Proposition 4.} {\sl The Moufang loop $G$ has a faithful representation if and only if it is isomorphically embedded into a loop of invertible elements of alternative algebra.}
\par\medskip
Now let $G$ be a simply connected analytical Moufang loop. In the neighborhood $U$ of the unity element the multiplication operation in $G$ is expressed through the addition and multiplication operations in the tangent Malcev algebra $A_{G}$ by the Campbell-Hausdorff series
\begin{eqnarray}
xy=x+y+\frac{1}{2}[xy]+\frac{1}{12}[[xy]y]+\frac{1}{12}[x[xy]]+\dots.
\end{eqnarray}
Define the mappings $(\rho,\lambda):U\to AutA_{G}$ by
\begin{eqnarray}
\rho_{x}=\sum^{\infty}_{n=0}{B_{n}\over n!} (\tau_{x})^{n},\qquad
\lambda_{x}=\sum^{\infty}_{n=0}{B_{n}\over n!} (-\tau_{x})^{n}
\end{eqnarray}
where $B_{n}$ are the Bernoulli numbers and $y\tau_{x}=[yx]$. Obviously, the operators $\rho_{x}$ and $\lambda_{x}$ are analytic, invertible, and satisfy the conditions of Proposition 1. Therefore the pair of mappings $(\rho,\lambda)$ is a representation of the local loop $U$. It is clear,
that this representation is faithful.
\par
Further, let $G\times A_{G}$ be a simply connected analytic Moufang loop which is locally isomorphic to the local loop $U\times A_{G}$. Obviously, $G$ is a subloop of $G\times A_{G}$. It follows from the left and right Moufang identities (2.1) that the relations
\begin{eqnarray}
\rho_{g_{i}g_{j}}=\lambda_{g_{i}}\rho_{g_{i}}\rho_{g_{j}}
\lambda_{g_{i}^{-1}},\quad
\lambda_{g_{j}g_{i}}=\rho_{g_{i}}\lambda_{g_{i}}\lambda_{g_{j}}
\rho_{g_{i}^{-1}}.
\end{eqnarray}
are valid for all $g_{i},g_{j}\in U$. Since the connected loop is algebraic generated by elements of any neighborhood of the unity element, we see that the multiplication in $G\times A_{G}$ can be represented as (3.1) where the elements $a$, $b$ are in $G$ and the operators $\rho_{b},\lambda_{a}$ are generated by the operators $\rho_{g_{i}}$, $\lambda_{g_{i}}$. Thus, we have the representation $(\rho,\lambda):G\to AutA_{G}$.
\par
Further, it follows from Proposition 2 and faithfully of representation of $U$ that the kernel $H=Ker(\rho,\lambda)$ is a discrete normal subloop of $G$. It follows from connectedness of $G$ that $H$ is in the center of $G$. Therefore there exists the representation $G\to AutF[H]$ regular on $H$ and identical on $G\setminus H$ (i.e. $a\rho_{b}=b\lambda_{a}=ab$ if $a,b\in H$ and $\rho_{a}=\lambda_{a}=id$ if $a,b\notin H$). Obviously, the induced representation $G\to Aut(A_{G}\otimes F[H])$ is faithful. Using Proposition 4, we get
\par\medskip
{\bf Theorem 1.} {\sl Every simply connected analytic Moufang loop is isomorphically embedded into a loop of invertible elements of alternative algebra over the field of real numbers.}
\par\medskip
Now let $G$ be an analytic Moufang loop locally isomorphic to the simply connected analytic Moufang loop $G'$ and suppose that the tangent algebra $A_{G}$ has the representation $\tau:A_{G}\to M$. It means that the split null extension $A_{G}\oplus M$ of $A_{G}$ is a Malcev algebra. Expressing the multiplication operation in the local Moufang loop $U\times M$ through the addition and multiplication operations in the tangent algebra $A_{G}\oplus M$, we'll have the representation $U\to AutM$ as was shown above. Continuing this representation from $U$ to $G'$, we'll receive the representation $G'\to AutM$.
\par
Conversely, let the analytical Moufang loop $G$ has the representation $G\to AutM$. This representation induces the representation $(\rho,\lambda):G'\to AutM$. We embed $G'$ into the loop $A^{*}$ of invertible elements of alternative algebra $A$. The enclosure $G'\to A^{*}$ induces the enclosure $\sigma:A_{G}\to A^{(-)}$. Suppose $\tau=\rho-\lambda$; then the mappings composition $\sigma\circ\tau:A_{G}\to EndM$ is a representation of $A_{G}$. Thus, we have the following
\par\medskip
{\bf Theorem 2.} {\sl Every $G$-module of the analytic Moufang loop $G$ is $A_{G}$-module of its tangent algebra. Conversely, every $A_{G}$-module of the Malcev algebra $A_{G}$ over the field of real numbers is $G'$-module of simply connected analytic Moufang loop $G'$ locally isomorphic to $G$.}
\par\medskip
Using Theorem 2 and Proposition 3, we get
\par\medskip
{\bf Theorem 3.} {\sl Every representation of semisimple analytic Moufang loop is completely reducible. Every irreducible $G$-module of analytic Moufang loop $G$ is either alternative or Malcev
irreducible bimodule.}

\section{Self-dual gauge fields in R$^{8}$}
Now we shall study the irreducible nonassociative representations of analytic Moufang loops. It follows from Theorem 3 that it is enough to investigate the regular berepresentation of complex and real Cayley-Dickson algebras and their commutator Malcev algebras.
\par
Suppose $A$ is a complex (real) Cayley-Dickson algebra, $M$ is its commutator
Malcev algebra, and ${\cal L}(A)$ is the enveloping Lie algebra of regular
representation of $A$. It is obvious that the algebra ${\cal L}(A)$ is
generated by the operators $R_{x}$ and $L_{x}$, where $x\in A$. We select in ${\cal L}(A)$ the subspaces $R(A)$, $L(A)$, $S(A)$, $P(A)$ and $D(A)$ generated by the operators $R_{x}$, $L_{x}$, $S_{x}=R_{x}+2L_{x}$, $P_{x}=L_{x}+2R_{x}$ and $D_{x,y}=[T_{x},T_{y}]+T_{[x,y]}$, where $T_{x}= R_{x}-L_{x}$, accordingly. Equations (3.2) imply that
\begin{eqnarray}
\left[R_{x},S_{y}\right]&=&R_{[x,y]},\\
\left[L_{x},P_{y}\right]&=&L_{[y,x]}.
\end{eqnarray}
It follows from the identities (4.1) and (4.2) that the algebra ${\cal L}(A)$ is decomposed into the direct sums
\begin{eqnarray}
{\cal L}(A)&=&D(A)\oplus S(A)\oplus R(A),\\
{\cal L}(A)&=&D(A)\oplus P(A)\oplus L(A),
\end{eqnarray}
of the Lie subalgebras $D(A)\oplus S(A)$, $D(A)\oplus P(A)$ and the vector spaces $R(A)$, $L(A)$ (See [15]). In addition, the map $x\to S_{x}$ from $M$ into $S(A)$ is a linear representation of the algebra $M$, which transforms the space $R(A)$ into $M$-module that is isomorphic, by (4.1), to the regular Malcev $M$-module. We prove the following
\par\medskip
{\bf Proposition 5.} {\sl The direct summands in (4.3) and (4.4) are orthogonal with respect to the scalar product $tr\{XY\}$ on ${\cal L}(A)$.}
\par\medskip
{\bf Proof.} Let $A$ be the complex Cayley-Dickson algebra. Then $A$ supposes the base $1,e_{1},...,e_{7}$ such that
$$
e_{i}e_{j}=-\delta_{ij}+c_{ijk}e_{k},
$$
where the structural constants $c_{ijk}$ are completely antisymmetric and different from 0 only if
$$
c_{123}=c_{145}=c_{167}=c_{246}=c_{257}=c_{374}=c_{365}=1.
$$
It is easy to see that in such base the operators
\begin{eqnarray}
R_{e_{i}}=e_{[i0]}-\frac12c_{ijk}e_{[jk]},
\nonumber\\
L_{e_{i}}=e_{[i0]}+\frac12c_{ijk}e_{[jk]},
\nonumber
\end{eqnarray}
where $e_{[\mu\nu]}$ are skew-symmetric matrixes $8\times 8$ with the elements $(e_{\mu\nu})_{\alpha\beta}=\delta_{\mu\alpha}\delta_{\nu\beta}-\delta_{\mu\beta}\delta_{\nu\alpha}$. Using the identity
$$
c_{ijk}c_{mnk}=\delta_{im}\delta_{jn}-\delta_{in}\delta_{jm}+c_{ijmn},
$$
where the completely antisymmetric tensor $c_{ijkl}$ is defined by the equality
$$
(e_{i},e_{j},e_{k})=2c_{ijkl}e_{l},
$$
we have
$$
D_{e_{i},e_{j}}=8e_{[ij]}+2c_{ijmn}e_{[mn]}.
$$
Finally, using the identities
\begin{eqnarray}
c_{mjk}c_{njk}=6\delta_{mn},
\nonumber\\
c_{ijmn}c_{kmn}=4c_{ijk},
\nonumber
\end{eqnarray}
we obtain
\begin{eqnarray}
tr\{R_{e_{i}}S_{e_{j}}\}=tr\{R_{e_{i}}D_{e_{j},e_{k}}\}=0,\\
tr\{L_{e_{i}}P_{e_{j}}\}=tr\{L_{e_{i}}D_{e_{j},e_{k}}\}=0.
\end{eqnarray}
As these equalities are invariant with respect to choice of the base in $A$, equalities (4.5) and (4.6) are valid for real forms of  the Lie algebra ${\cal L}(A)$. This completes the proof of Proposition 5.
\par
Let $A$ be the real Cayley-Dickson algebra with division (octonion algebra). Its enveloping Lie algebra ${\cal L}(A)$ (in fixed base) consists of real skew-symmetric $8\times8$ matrixes. Therefore we can connect every element $F=F_{\mu\nu}e_{[\mu\nu]}$ of ${\cal L}(A)$ with the 2-form
$F=F_{\mu\nu}dx^{\mu}\wedge dx^{\nu}$. It follow from (4.5) and (4.6) that the factors of $F$ are such that
\begin{eqnarray}
\varepsilon F_{0i}+\frac12c_{ijk}F_{jk}=0, &\qquad &\mbox{if} \qquad
\left\{
\begin{array}{l}
F\in S(A)\oplus D(A)\\
F\in P(A)\oplus D(A)\\
\end{array},
\right.\\
\varepsilon F_{0i}=c_{ijk}F_{jk}, &\qquad &\mbox{if} \qquad
\left\{
\begin{array}{l}
F\in R(A)\\
F\in L(A)\\
\end{array},
\right.
\end{eqnarray}
where there is no summing over $j,k$ in (4.8), $c_{ijk}\ne 0$, and
\par
$\varepsilon=0$, if $F\in D(A)$,
\par
$\varepsilon=1$, if $F\in S(A)\oplus D(A)$, $F\notin D(A)$ or $F\in R(A)$,
\par
$\varepsilon=-1$, if $F\in P(A)\oplus D(A)$, $F\notin D(A)$ or $F\in L(A)$.
\par\noindent
For $\varepsilon=-1$ these are precisely the (anty-self-dual) equations of Corrigan et al. [4] It follows from Proposition 5 that the expressions (4.7) and (4.8) are not depend with respect to choice of base in $A$. In addition, $R (A)$ and $L(A)$ are not Lie algebras. Therefore the equations (4.8), in contrast to (4.7), are not Yang-Mills equations.
\par
Nevertheless, there are a solution of the equations (4.8), which generalizes the known (anty-)instanton solution of Belavin et al.~[1] We find this solution. For this purpose we consider ${\cal L}(A)$-valued 1-forms
\begin{eqnarray}
A_{\lambda, b}(x)&=&\frac12 \frac{R_{x-b}R_{\partial_{\nu}\bar x}-R_{\partial_{\nu}x} R_{\bar x-\bar b}}{\lambda^{2}+|x-b|^{2}}dx^{\nu},
\nonumber\\
B_{\lambda, b}(x)&=&\frac12 \frac{L_{x-b}L_{\partial_{\nu}\bar x}-L_{\partial_{\nu}x} L_{\bar x-\bar b}}{\lambda^{2}+|x-b|^{2}}dx^{\nu},
\nonumber
\end{eqnarray}
where $x,b\in {A}$ and $\lambda$ is a positive number. The following proposition is true.
\par\medskip
{\bf Proposition 6.} {\sl There exists a smooth pair of mappings $(f,g):A\to{\bf S}^{7}$,
determined near $\infty$, such that $A_{\lambda,b}(x)\sim f^{-1}(x)\,df(x)$ and $B_{\lambda,b}(x)\sim g^{-1}(x)\,dg(x)$ as $|x |\to\infty$ for any $\lambda$ and $b$. The difference $F_{\lambda,b}-G_{\lambda,b}$ of corresponding 2-form $F_{\lambda,b}=dA_{\lambda,b}+A_{\lambda,b}\wedge A_{\lambda,b}$ and $G_{\lambda,b}=dB_{\lambda,b}+B_{\lambda,b}\wedge B_{\lambda,b}$ is a sum of self-dual and anty-self-dual 2-forms.}
\par\medskip
{ \bf Proof.} Using the identities
$$
2\partial_{\nu}\frac{x}{|x|}=\left(\frac{\partial_{\nu}\bar x}{|x|}-\frac{\bar x(\partial_{\nu}x)\bar x}{|x|^{3}}\right)+\left(\frac{\partial_{\nu}\bar x}{|x|^{3}}+\frac{\bar x(\partial_{\nu}x)\bar x}{|x|^{5}}\right),
$$
\begin{eqnarray}
R_{x}R_{\bar x}=|x|^{2}id,& \qquad R_{x}R_{\partial_{\nu}\bar x}R_{x}=R_{x(\partial_{\nu}\bar x)x},&
\nonumber \\
L_{x}L_{\bar x}=|x|^{2}id,& \qquad L_{x}L_{\partial_{\nu}\bar x}L_{x}=L_{x(\partial_{\nu}\bar x)x},&
\nonumber
\end{eqnarray}
we get
\begin{eqnarray}
A_{\lambda,b}(x)&\sim& f^{-1}(x)\,df(x),
\nonumber\\
B_{\lambda,b}(x)&\sim& g^{-1}(x)\,dg(x),
\nonumber
\end{eqnarray}
where $f(x)=R_{\bar x}|x|^{-1}$, $g(x)=L_{\bar x}|x|^{-1}$, and
\begin{eqnarray}
F_{\lambda,b}&=&\lambda^{2}\frac{R_{\partial_{\mu}x}R_{\partial_{\nu}\bar
x}-R_{\partial_{\nu}x} R_{\partial_{\mu}\bar x}}{(\lambda^{2}+|x-b|^{2})^{2}}dx^{\mu}\wedge dx^{\nu},
\nonumber\\
G_{\lambda,b}&=&\lambda^{2}\frac{L_{\partial_{\mu}x}L_{\partial_{\nu}\bar
x}-L_{\partial_{\nu}x} L_{\partial_{\mu}\bar x}}{(\lambda^{2}+|x-b|^{2})^{2}}dx^{\mu}\wedge dx^{\nu}.
\nonumber
\end{eqnarray}
Let $x=x_{0}+{\bf x}$ and $\bar x=x_{0}-{\bf x}$. Then, by (4.1) and (4.1), we get that $F_{\lambda,b}-G_{\lambda,b}=R_{\lambda,b}+L_{\lambda,b}$, where
\begin{eqnarray}
R_{\lambda,b}&=&\frac{-2\lambda^{2}R_{e_{i}}}{(\lambda^{2}+|x-b|^{2})^{2}}(dx^{0}\wedge x^{i}+c_{ijk}dx^{j}\wedge dx^{k}),
\nonumber\\
L_{\lambda,b}&=&\frac{-2\lambda^{2}L_{e_{i}}}{(\lambda^{2}+|x-b|^{2})^{2}}(dx^{0}\wedge x^{i}-c_{ijk}dx^{j}\wedge dx^{k}).
\nonumber
\end{eqnarray}
The self-dual of $R_{\lambda,b}$ and  anty-self-dual of $L_{\lambda,b}$ are
obvious. To conclude the proof, it remains to note that the map $f$, $g$ translates ${\bf S}^{7}$ in itself and are the symmetries $x\to R_{\bar x}$, $x\to L_{\bar x}$ on ${\bf S}^{7}$.

\small
\newpage
\bigskip
\begin{center}
{\bf References}
\end{center}
\medskip
\small
\begin{itemize}
\item[[1{]}]
A.A.~Belavin, A.S.~Polyakov, A.S.~Schwarz, and Yu.S.~Tyupkin, Phys. Lett. B59 (1975) 85.
\item[[2{]}]
E.~Cremmer, B.~Julia, and J.~Scherk, Phys. Lett. B76 (1978) 409;
E.~Witten, Nucl. Phys. B186 (1981) 412.
\item[[3{]}]
F.~Englert, Phys. Lett. B119 (1982) 339;
R.D'Auria, P.~Fre, and P.~Van Nieuwenhuizen, Phys. Lett. B122 (1983) 22;
F.~Gursey and C.-H.~Tze, Phys. Lett. B127 (1983) 22.
\item[[4{]}]
E.~Corrigan, C.~Devchand, D.B.~Fairlie and J.~Nuyts, Nucl. Phys. B 214 (1983) 452.
\item[[5{]}]
B.~Crossman, T.W.~Kephart, J.D.~Stasheff, Comm. Math. Phys. 96 (1984) 431;
D.B.~Fairlie, J.~Nuyts, J. Phys. A17 (1984) 2867;
S.~Fubini, H.~Nicolai, Phys. Lett. B155 (1985) 369;
A.H.~Bilge, T.~Dereli, S.~Kocak, J. Math. Phys. 38 (1997) 4804;
E.G.~Floratos and A.~Kehagias, Phys. Lett. B427 (1998) 283.
\item[[6{]}]
T.A.~Ivanova, Phys. Lett. B315 (1993) 277;
M.~Gunaydin, H.~Nicolai, Phys. Lett. B351 (1995) 169;
E.G.~Floratos, G.K.~Leontaris, A.P.~Polychronakos, and R.~Tzani, Phys. Lett. B421 (1998) 125.
\item[[7{]}]
J.A.~Harvey, A.~Strominger, Phys. Rev. Lett. 66 (1991) 549;
E.G.~Floratos and G.K.~Leontaris, Nucl. Phys. B512 (1998) 445.
\item[[8{]}]
B.S.~Acharya, M.~O'Loughlin, B.~Spence, Nucl. Phys. B503 (1997) 657;
L.~Baulieu, H.~Kanno, I.M.~Singer, Commun. Math. Phys. 194 (1998) 149.
\item[[9{]}]
S.K.~Donaldson, Topology 29 (1990) 257;
E.~Witten, Commun. Math. Phys. 117 (1988) 353.
\item[[10{]}]
T.~Curtright, D.B.~Fairlie, C.K.~Zachos, Phys. Lett. B405 (1997) 37;
J.M.~Figueroa-O'Farrill, C.~Kohl, B.~Spence, Nucl. Phys. B521 (1998)  419;
C.M.~Hull, Adv. Their. Math. Phys. 2 (1998) 619.
\item[[11{]}]
R.~Dundarer, F.~Gursey, and C.-H.~Tze, J. Math. Phys. 25 (1984) 1496.
\item[[12{]}]
J.R.~Faulkner and J.C.~Far, Bull. London Math. Soc. 9 (1977) 1.
\item[[13{]}]
J.~Lohmus, E.~Paal, and L.~Sorgsepp, Nonassociative Algebras in Physics
(Hadro\-nic Press, Inc., Palm Harbor, 1994).
\item[[14{]}]
R.H.~Bruck, A Survey of Binary System
(Berlin-Heidelberg-New York:Sprin\-ger-Verlag, 1971).
\item[[15{]}]
R.D.~Schafer, An Introduction to Non-Assocoative Algebras
(Academic, New York, 1966).
\item[[16{]}]
K.A.~Zhevlakov, A.M.~Slinko, I.P.~Shestakov, and A.I.~Shirsov,
Rings that are nearly associative (Academic Press, New York, 1982).
\item[[17{]}]
A.A.~Sagle, Trans. Am. Math. Soc. 101 (1961) 426.
\item[[18{]}]
A.A~Sagle, Pacif. J. Math. 12 (1962) 1057;
V.T.~Filippov, Algebra i Logika (Russian) 15 (1976) 235.
\item[[19{]}]
E.N.~Kuzmin Algebra i Logika (Russian) 7 (1968) 48.
\item[[20{]}]
A.~Elduque, Proc. Amer. Math. Soc. 107 (1989) 73.
\item[[21{]}]
A.I.~Malcev, Math. Sb. (Russian) 36 (1955) 569;
E.N.~Kuzmin, Algebra i Logika (Russian) 10 (1971) 3.
\item[[22{]}]
F.S.~Kerdman, Algebra i Logika (Russian) 18 (1979) 523.
\item[[23{]}]
S.~Eilenberg, Ann. Soc. Polon. Mat. {\bf 21} (1948) 125;
N.~Jacobson, Structure and Representations of Jordan Algebras (Providence, R. I., 1968).
\item[[24{]}]
R.D.~Schafer, Trans. Amer. Math. Soc. 72 (1952) 1;
K.~McCrimmon, Proc. Amer. Math. Soc. 18 (1966) 480.
\item[[25{]}]
A.~Elduque, Comm. Algebra 18 (1990) 1551.
\item[[26{]}]
E.K.~Loginov, Comm. Algebra 21 (1993) 2527.
\end{itemize}

\end{document}